\documentclass[traditabstract]{aa}
\usepackage{longtable}
\usepackage[pdftex]{graphicx}
\usepackage{times}                          
\usepackage{graphicx}
\usepackage{rotating}
\usepackage{multirow}
\usepackage{amssymb}
\usepackage{amsmath}
\usepackage{amstext}
\usepackage{listings}
\usepackage{lscape}
\usepackage{txfonts}
\usepackage{float}
\usepackage{natbib}
\bibpunct{(}{)}{;}{a}{}{,}
\usepackage{color}
\usepackage{longtable}

\begin{document}

\title{A study of the association of \textit{Fermi} sources with massive young galactic objects}

\author{P.~Munar-Adrover\inst{\ref{inst1}}
\and J. M.~Paredes\inst{\ref{inst1}}
\and G. E.~Romero{\inst{\ref{inst2}}\inst{,\ref{inst3}}}}

\institute{Departament d'Astronomia i Meteorologia, Institut de Ci\`encies del Cosmos (ICC), Universitat de Barcelona (IEEC-UB), Mart\'{\i} i Franqu\`es 1, 08028 Barcelona, Spain  \email{pmunar@am.ub.es}\label{inst1}
\and
Instituto Argentino de Radioastronom\'ia, CONICET, C.C.5, Villa Elisa, 1894, Argentina \email{romero@iar.unlp.edu.ar}\label{inst2}
\and
Facultat de Ciencias Astron\'omicas y Geof\'isicas, Universidad Nacional de La Plata, Paseo del Bosque, La Plata, 1900, Argentina\label{inst3}}

\date{Received 25 January 2011 /
Accepted ...}

\abstract{Massive protostars have associated bipolar outflows that can produce strong shocks when they interact with the surrounding medium. At these shocks particles can be accelerated up to relativistic energies. Relativistic electrons and protons can then produce gamma-ray emission, as some theoretical models predict. To identify young galactic objects that may emit gamma rays, we crossed the \textit{Fermi} First Year Catalog with some catalogs of known massive young stellar objects (MYSOs), early type stars, and OB associations, and we implemented Monte Carlo simulations to find the probability of chance coincidences. We obtained a list of massive MYSOs that are spatially coincident with \textit{Fermi} sources. Our results indicate that $\sim 70\%$ of these candidates should be gamma-ray sources with a confidence of $\sim5\sigma$. We studied the coincidences one by one to check the viability of these young sources as potential counterparts to \textit{Fermi} sources and made a short list of best targets for new detailed multifrequency observations. The results for other type of young galactic objects are not conclusive.}


\maketitle

\section{Introduction}

Recently, massive young stellar objects (MYSOs) have been suggested as gamma-ray sources \citep{ara07,rome08,bosch10}. Massive stars are formed in dense cores of cold clouds. The processes that take place during the formation of the star are mostly unknown. It is clear, however, that the formation of massive stars involves outflows \citep{gara99, reip01}. The accumulation of material around the core of the cloud would generate a massive protostar that starts to accrete material from the environment. The accretion is expected to have angular momentum that leads to the formation of an accretion disk. The rotation would twist the strong magnetic fields present in the progenitor cloud around the disk, where a magnetic tower can be formed, giving rise to collimated outflows or jets, as simulations predict \citep{baner06, baner07}. The observational evidence of outflows comes from methanol masers and from direct detection of thermal radio jets. These jets propagate along distances in a fraction of a parsec \citep{mar93}. At the jet termination region, interaction with the external medium creates two shocks: a bow shock moving in the interstellar medium (ISM) and a reverse shock in the jet. These shocks can accelerate particles that, in turn, can produce gamma rays trough inverse Compton (IC) scattering of infrared (IR) photons, relativistic Bremsstrahlung or inelastic proton-proton collisions, if protons are accelerated as well. In some cases non~thermal radio lobes and jets have been observed, indicating the presence of relativistic electrons that produce synchrotron radiation \citep{gara03, carrasco10}.\\

Other possible scenarios have been suggested for the gamma-ray production involving young stars, such as the case of the massive stars with strong winds. In this scenario, gamma-rays could be produced in the interaction between the supersonic winds and the ISM. The terminal shock can accelerate particles and ions up to high energies, which might interact with the ambient matter producing gamma-rays. Whereas the luminosity produced by a single massive star wind should be low, collective effects might be important \citep{torr04}. \\
Gamma-rays can also be produced in the wind interaction region of a WR+OB binary system \citep{bena03}. In this case, the acceleration region is between the two components of the binary system and is exposed to strong photon fields where IC cooling of the electrons can generate a significant amount of high-energy (HE) non~thermal emission. A source of this class, $\rm{\eta-Carinae}$, has been recently detected at $E > 100$ MeV by \textit{AGILE} \citep{tava09}.

Of-type stars have strong winds with velocities higher than the escape velocity, which implies a strong mass loss rate ($10^{-6}-10^{-5}$ M$_{\odot}$ yr$^{-1}$). The action of this wind in the interstellar medium can create hot gas bubbles with expanding boundaries of swept-up material, which might produce gamma rays in a similar way to the case of WR stars. The case of gamma-ray emission in Of-type stars has been discussed in the past, e.g. by \cite{volk82}. The predicted luminosity, however, is still below the current sensitivity of gamma-ray instruments.\\

Finally, OB associations are tracers of a number of galactic objects that can produce gamma rays, such as neutron stars, massive stars with strong winds, young stellar objects (YSOs), etc. They are also thought to be places where acceleration of a significant fraction of galactic cosmic rays (CRs) might occur \citep[e.g.][]{binns08}. \\

The aim of this work is to find evidence supporting the presence of HE emission coming from massive YSOs and other young galactic sources. To attain this goal we study the spatial coincidence between gamma-ray sources detected by \textit{Fermi} and samples of young objects, such as YSOs, WR stars, Of-type stars, and OB associations. We also estimate the probability of chance coincidences by using Monte Carlo simulations, and we provide a list of counterpart candidates of the gamma-ray sources.

\section{Cross-correlation of the First \textit{Fermi} Catalog with massive young galactic objects}

There is not observational evidence of any YSO emitting gamma-rays so far. The new generation of gamma-ray telescopes, like \textit{Fermi}, will make it possible to reach the necessary sensitivity level to detect these faint gamma-ray sources soon, in case the predictions were right. To identify those young objects that might be emitting gamma rays, we first took the recently published First \textit{Fermi} Catalog \citep{abdo10} by the \textit{Fermi} Collaboration and we excluded all known firm identifications, getting a list of 1392 sources. Then we crossed this list with catalogs of confirmed and well characterized YSOs and other type of young stars. We also did a Monte Carlo study to determine the probability of pure chance coincidences between the crossed catalogs. 

\subsection{Catalogs}

The catalogs used in this study are listed in Table \ref{tab:catalogs}. Here we describe each one in more detail.

The \textit{Fermi Large Area Telescope} First Catalog \citep{abdo10} contains the detected sources during the first 11 months of the science phase of the mission, which began on 2008 August 4. This catalog contains 1451 gamma-ray sources detected and characterized in the 100 MeV to 100 GeV range with a typical position uncertainty of $\sim 6^\prime$. After excluding the firm identifications from the original sample, we get 1392 sources. Most of them are located on the Galactic plane (see Figure \ref{fig:hist}). 
\\

The Red \textit{MSX} Source (RMS) survey is an ongoing multi~wavelength (from radio to infrared) observational program with the objective of providing a well-selected sample of MYSOs in the entire Galaxy \citep{urqu08}. About $\sim$2000 MYSO candidates have been identified by comparing the colors of \textit{MSX} and 2MASS point sources (at 8, 12, 14, and 23 $\mu$m) with those of well known MYSOs. The survey also uses high-resolution radio continuum observations at 6~cm obtained with the VLA in the northern hemisphere and at 3.6~cm and 6cm with ATCA in the southern hemisphere. They help to distinguish between genuine MYSOs and other types of objects, such as ultracompact HII regions, evolved stars, or planetary nebulae, which contaminate the sample. In addition to these targeted observations, archival data of a previous VLA survey of the inner Galaxy were used. This ongoing program has provided a sample of 637 well-identified MYSOs until now, which were used in our work.
\\

The VIIth catalog of Population I WR stars \citep{vander01} contains 227 stars, with spectral types and \textit{bv} photometry. In recent years, the number of WR stars has increased in 71 new stars, respect to the VIth catalog and the coordinates have also been improved. The position uncertainty is close to a fraction of an arcsecond. 
\\

The catalog of Of-type stars is the one of \cite{cruz74}, which contains 664 stars. The catalog provides $\rm{m_v}$, B--V, spectral type, radial velocity, radial component of the peculiar velocity, possible multiplicity of the object, and other characteristics for each source. The typical uncertainty in the star position is $\sim 1^\prime$.
\\

Finally, the catalog of OB associations is the one by \cite{melnik95}. This catalog contains 88 associations and provides distances to the association, number of stars, and size of the association along the Galactic latitude and longitude axes. The typical value of the size is $\sim 20-30$~pc.

\subsection{Spatial coincidences}

We crossed the \textit{Fermi} catalog with the catalogs of young galactic objects mentioned above. We calculated the distance between two sources using the statistical parameter $S$ \citep{aling82}:
\\
\begin{equation*} 
S=\sqrt{\frac{(\Delta \alpha \cos \delta)^2}{\sigma_{i_\alpha}^2+\sigma_{j_\alpha}^2} + \frac{\Delta \delta^2}{\sigma_{i_\delta}^2+\sigma_{j_\delta}^2}}
\end{equation*}
\\
where $\Delta\alpha$ and $\Delta\delta$ are the difference between the right ascension and the declination of the two compared sources, respectively, $\sigma_{a_b}$ is the uncertainty in the position of the source, and $(i,j)$ represent the two sources. The error in the position of the \textit{Fermi} sources is taken as the 95$\%$ confidence ellipse. The error in the position of the other compared sources is the precision in the coordinates for YSOs, WR stars and Of-type stars, and the angular size of the association in the case of OB associations. If $S$ is lower than or equal to the unit, it means that the source position (YSO, WR, Of-type, or OB associations) is inside the 95$\%$ uncertainty ellipse of the \textit{Fermi} source, within its own position uncertainty, and that case is considered to be a coincidence. Massive YSO and protostar are point-like objects compared with the confidence contours of Fermi sources. The same is valid for WR and Of stars, either in binary systems or isolated. OB associations are large  systems that can contain more than a single gamma-ray source. We note that our study is based on two-dimensional coincidences, since we are comparing the equatorial coordinates of the sources.

\subsection{Monte Carlo analysis}{\label{subs:monte}}

\begin{figure}
\resizebox{\hsize}{!}{\includegraphics{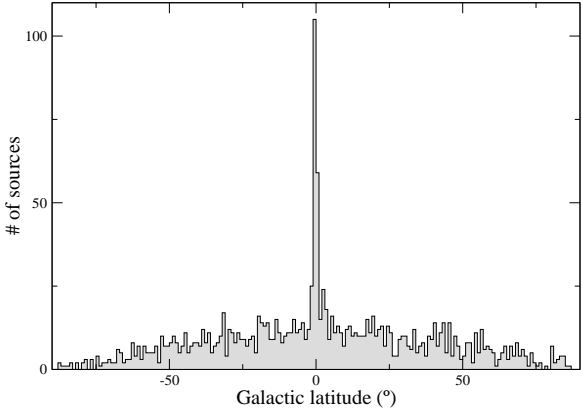}}
  \caption{Distribution of First \textit{Fermi} Catalog sources in galactic latitude. \label{fig:hist}}
\end{figure}

To determine the chance coincidences we used the Monte Carlo method for simulating sets of synthetic gamma-ray sources starting from the \textit{Fermi} Large Area Telescope First Catalog. We followed a similar criteria to that used by \cite{rome99} to search for the possible association of unidentified \textit{EGRET} sources with other type of celestial objects. In this algorithm, the galactic coordinates of a gamma-ray source $\left( l, b \right)$ are moved to new ones $\left( l^\prime, b^\prime \right)$. The new galactic longitude coordinate is calculated by doing $l^\prime=l + R_1 \times 360^\circ$, where $R_1$ is a random number between 0 and 1 that never repeats from source to source or from set to set. Since the distribution of \textit{Fermi} sources is almost constant in Galactic longitude, we do not impose any constraint on this coordinate in the simulations. The sources in the \textit{Fermi} catalog have a given distribution in galactic latitude (see Figure \ref{fig:hist}). In order to constrain the simulations with this distribution, the galactic latitude coordinate is calculated by doing $b^\prime=b + R_2\times 1^\circ$, where again $R_2$ is a random number between 0 and 1. Here, if the integer part of $b^\prime$ is greater than the integer part of $b$ or the sign of $b^\prime$ is different than the sign of $b$, then $b^\prime$ is replaced by $b^\prime-1^\circ$.

\begin{table}[htb]
  \begin{center}
    \caption{List of the catalogs used in the study.\label{tab:catalogs}}        
    \scalebox{0.90}[0.90]{
    \begin{tabular}{lll}
    \hline
    \hline
     Object & Catalog & $\#$ of \\
	 type   &         & sources \\
     \hline
     $\gamma$-ray sources & First \textit{Fermi} Catalog\tablefootmark{a} & 1392\tablefootmark{f}\\
     YSO & RMS Survey\tablefootmark{b}& 637\\
     WR & VII$^{th}$ Catalog of Galactic Wolf-Rayet stars\tablefootmark{c} &  227\\ 
     Of-type & A catalog of galactic O stars and&  664 \\
	    & the ionization of the low density & \\
	    &  interstellar medium by runaway stars \tablefootmark{d} & \\
     OB associations & A new list of OB & 88\\
        & associations in our Galaxy\tablefootmark{e}& \\
     \hline
     \end{tabular}}
     \tablefoot{\tablefoottext{a}{\cite{abdo10}},
	 \tablefoottext{b}{\cite{urqu08}},
	 \tablefoottext{c}{\cite{vander01}},
	 \tablefoottext{d}{\cite{cruz74}},
	 \tablefoottext{e}{\cite{melnik95}};
     \tablefoottext{f}{number of gamma-ray sources after excluding the firm identifications}}
  \end{center}
\end{table} 

We simulated 1500 sets of synthetic \textit{Fermi} sources and each set was compared with a fixed set of different kinds of objects: YSOs, WR stars, Of-type, and OB associations. In each simulation, we calculated the distance between the two sources using the statistical parameter \emph{S}.

For each kind of compared objects we calculated the average number of coincidences and its standard deviation after all Monte Carlo simulations. We calculated the chance coincidence probability using the actual number of coincidences for each type of object and assuming a Gaussian distribution in the simulations. This probability allows us to know the reliability of our study. 
We also repeated this process moving the \textit{Fermi} sources in $2^\circ$-bins in galactic latitude, i.e. replacing the galactic latitude coordinate by $b^\prime=b + R_2\times 2^\circ$. If the integer part of $b^\prime$ is greater than the integer part of $b$ or the sign of $b^\prime$ is different than the sign of $b$, then $b^\prime$ is replaced by $b^\prime-2^\circ$. This binning allows us to keep the initial distribution in galactic latitude as well.

\begin{table}[htb]
  \begin{center}
    \caption{Statistical results obtained from simulations. Latitude galactic coordinate has been constrained, while galactic longitude remains free.\label{tab:stat}}
    \scalebox{0.80}[0.80]{
    \begin{tabular}{llr@{$\pm$}llr@{$\pm$}ll}
    \hline
    \hline
Object&Coincident &\multicolumn{2}{c}{Simulated} &Probability&\multicolumn{2}{c}{Simulated} &Probability\\
type&$\gamma$-ray sources &\multicolumn{2}{c}{$1^\circ -bin$} &$1^\circ -bin$&\multicolumn{2}{c}{$2^\circ -bin$} &$2^\circ -bin$\\
\hline
YSO & 12 & 4.4	& 2.0 & 1.8$\times 10^{-4}$ & 3.6 & 1.8 & 5.6$\times 10^{-6}$\\
WR & 2 & 1.3 & 1.1 & 2.9$\times 10^{-1}$ & 1.2 & 1.1 & 2.9$\times 10^{-1}$\\
Of-type & 5 & 2.9 & 1.7 & 1.1$\times 10^{-1}$ & 2.9 & 1.7& 1.1$\times 10^{-1}$\\
OB assoc. & 107 & 72.5& 8.0& 4.2$\times 10^{-6}$& 72.8&8.0&5.5$\times 10^{-6}$\\
 \hline
 \end{tabular}}
  \end{center}
\end{table} 

%
\begin{table*}[h!]
  \begin{center}
    \caption{Positional coincidence of \textit{Fermi} sources with MYSOs. \label{tab:yso}}    
      \scalebox{0.692}[0.692]{
    \begin{tabular}{llllr@{$\pm$}lr@{$\pm$}llllllll}
    \hline
    \hline
Fermi Name&RA&Dec&95\% Semi& \multicolumn{2}{c}{Spectral Index $\Gamma$}& \multicolumn{2}{c}{Flux(E$>$100 MeV)}&MSX Name&RA&Dec&$\Delta\theta$&Distance\tablefootmark{*}&$L_{bol}$\tablefootmark{*}& Mass\\
(1FGL)&$\left(^\circ \right)$&$\left(^\circ \right)$&Major Axis ($^\circ$)& \multicolumn{2}{c}{$(F \propto E^{-\Gamma})$}& \multicolumn{2}{c}{$\times 10^{-11}$ erg cm$^{-2}$ s$^{-1}$}&&$\left(^\circ \right)$&$\left(^\circ \right)$&$\left(^\circ \right)$&$\left(\rm{kpc}\right)$&$\left( \times 10^3 \rm{L}_{\odot}\right)$&$\left( \rm{M}_\odot \right)$\\
\hline
J0541.1$+$3542&85.2805&35.7091&0.1397&2.41&0.13&1.6&0.5&G173.6328$+$02.8064&85.27929&$+$35.82633&0.12&1.6\tablefootmark{a}&$4.8$\tablefootmark{d}&\\
				     &&&&\multicolumn{2}{c}{}&\multicolumn{2}{c}{}&G173.6339$+$02.8218&85.29592&$+$35.83380&0.13&1.6\tablefootmark{a}&3.2\tablefootmark{e}&\\
				     &&&&\multicolumn{2}{c}{}&\multicolumn{2}{c}{}&G173.6882$+$02.7222&85.22758&$+$35.73558&0.05&1.6\tablefootmark{a}&--&\\
\\				     
J0647.3$+$0031&101.8417&0.5289&0.2150&2.41&0.11&1.9&0.5&G212.0641$-$00.7395&101.80567&$+$0.43514&0.10&6.4\tablefootmark{b}&25\tablefootmark{f}&\\
\\
J1256.9$-$6337&194.2474&$-$63.6212&0.1955&2.26&0.12&4.9&1.1&G303.5990$-$00.6524&194.35546&$-$63.51650&0.12&11.3\tablefootmark{b}&8.3\tablefootmark{f}&\\
\\
J1315.0$-$6235&198.7635&$-$62.5971&0.1860&2.31&0.12&6.9&0.0&G305.4840$+$00.2248&198.40016&$-$62.53708&0.18&3.6\tablefootmark{b}&3.8\tablefootmark{f}&\\
\\
J1651.5$-$4602&252.8831&$-$46.0340&0.2258&2.21&0.07&13.9&3.4&G339.8838$-$01.2588\tablefootmark{1}&253.01942&$-$46.14267&0.14&2.6\tablefootmark{b}&21.0\tablefootmark{f}&\\
\\
J1702.4$-$4147&255.6039&$-$41.7859&0.0800&2.39&0.07&8.7&2.0&G344.4257$+$00.0451B&255.53674&$-$41.78303&0.05&5.0\tablefootmark{b}&15.0\tablefootmark{f}&\\
				     &&&&\multicolumn{2}{c}{}&\multicolumn{2}{c}{}&G344.4257$+$00.0451C&255.53587&$-$41.78617&0.05&5.0\tablefootmark{b}&15.0\tablefootmark{f}&\\
\\
J1846.8$-$0233&281.7001&$-$2.5628&0.1262&2.21&0.06&9.3&2.3&G030.1981$-$00.1691&281.76274&$-$2.51003&0.08&7.4\tablefootmark{b}&29.0\tablefootmark{f}&\\
\\
J1848.1$-$0145&282.0470&$-$1.7605&0.0859&2.23&0.04&9.5&3.2&G030.9726$-$00.1410&282.09178&$-$1.80842&0.07&5.7\tablefootmark{b}&3.9\tablefootmark{f}&1.9$\times 10^3$\tablefootmark{g}\\
		   		     &&&&\multicolumn{2}{c}{}&\multicolumn{2}{c}{}&G030.9959$-$00.0771&282.04516&$-$1,75808&0.0044&5.7\tablefootmark{b}&5.1\tablefootmark{f}&1.9$\times 10^3$\tablefootmark{g}\\
\\
J1853.1$+$0032&283.2884&0.5366&0.5207&2.18&0.07&5.7&1.7&G032.8205$-$00.3300&282.04436&$-$1.75703&0.34&5.1\tablefootmark{b}&17.0\tablefootmark{f}&\\
				     &&&&\multicolumn{2}{c}{}&\multicolumn{2}{c}{}&G033.3891$+$00.1989&282.89092&$+$0.49750&0.40&5.1\tablefootmark{b}&11.0\tablefootmark{f}&\\
				     &&&&\multicolumn{2}{c}{}&\multicolumn{2}{c}{}&G033.3933$+$00.0100&283.06109&$+$0.41528&0.26&6.8\tablefootmark{b}&7.9\tablefootmark{e}&\\				     
					 &&&&\multicolumn{2}{c}{}&\multicolumn{2}{c}{}&G033.5237$+$00.0198 & 283.11179 & $+$0.53569 & 0.34 & 6.8\tablefootmark{b}&7.9\tablefootmark{e}& \\				    				  				  
				     &&&&\multicolumn{2}{c}{}&\multicolumn{2}{c}{}&G034.0126$-$00.2832&283.60437&$+$0.83239&0.43&13.3\tablefootmark{b}&34.0\tablefootmark{f}&\\
				     &&&&\multicolumn{2}{c}{}&\multicolumn{2}{c}{}&G034.0500$-$00.2977&283.63454&$+$0.85914&0.47&13.3\tablefootmark{b}&24.0\tablefootmark{f}&\\
\\
J1925.0$+$1720&291.2748&17.3485&0.1443&2.28&0.12&2.4&1.0 &G052.2025$+$00.7217A&291.24933&$+$17.42169&0.08&10.2\tablefootmark{b}&15.0\tablefootmark{f}&\\
					 &&&&\multicolumn{2}{c}{}&\multicolumn{2}{c}{}&G052.2078$+$00.6890&291.28553&$+$17.41317&0.07&10.2\tablefootmark{b}&20.0\tablefootmark{f}&\\
\\				     
J1943.4$+$2340&295.8667&23.6815&0.1118&2.23&0.11&2.6&0.7&G059.7831$+$00.0648\tablefootmark{2,3}&295.79680&$+$23.73433&0.08&2.2\tablefootmark{a}&6.8\tablefootmark{f}&840 and 190 \tablefootmark{h}\\
\\
J2040.0$+$4157&310.0154&41.9533&0.1970&2.66&0.06&7.9&1.2&G081.5168$+$00.1926&309.99066&$+$41.98739&0.04&1.7\tablefootmark{c}&0.704\tablefootmark{f}&\\
\\
 \hline
 \end{tabular}}
      \tablefoot{\tablefoottext{1}{Detected in radio at 8.6 GHz with integrated flux of 2.6 mJy; }
      			 \tablefoottext{2}{Detected in radio at 8.6 GHz with integrated flux of 1.0 mJy. }
      			 \tablefoottext{3}{Observed in the X-ray with \textit{Chandra} \citep{beut02}; }      
      			Distances: 
      						\tablefoottext{a}{distance to the complex, taken from the literature, }
      						\tablefoottext{b}{kinematic distance determined from the systemic velocity of the complex, }
      						\tablefoottext{c}{distance has been taken from the literature. }
      			Luminosities:			
      						\tablefoottext{d}{\textit{IRAS} fluxes, }
      						\tablefoottext{e}{\textit{MSX} 21 $\mu$m band flux using a scaling relationship determined from a comparison with sources where spectral energy distribution (SED) 													   fits have been possible. }
      						\tablefoottext{f}{SED fit to the available infrared fluxes (\textit{2MASS, MSX, MIPSGAL/IGA}) and literature (sub)millimetre fluxes.}
      						\tablefoottext{*}{data obtained from  $\rm{http://www.ast.leeds.ac.uk/cgi-bin/RMS/RMS\_DATABASE.cgi}$. }
				Masses:      						
      						\tablefoottext{g}{from \cite{ragan06}, }
      						\tablefoottext{h}{mass of the 2 cores identified in \cite{beuther02b}}}
  \end{center}
\end{table*}

\section{Results}{\label{sec:results}}

The results from our statistical study are shown in Table \ref{tab:stat}. In this table we list, from left to right, the object type, the number of coincidences between the each compared catalog and the original \textit{Fermi} catalog, the simulated average number of coincidences, and the chance coincidence probability for each binning in galactic latitude. We find 12 gamma-ray sources spatially coincident with YSOs, 2 with WR stars, 5 with Of-type stars and 107 with OB associations. 

From the Monte Carlo analysis, we see that there is a strong correlation between gamma-ray sources and YSOs: the catalog cross-check returns 12 coincidences between gamma-ray sources and YSOs. The Monte Carlo simulations, for the case of displacing the \textit{Fermi} sources in 1$^\circ$-bins, returns an average of coincidences of 4.4$\pm$2.2 sources, which is the number of chance coincidences. This means that 7.6 of the 12 coincident \textit{Fermi} sources ($\sim 63\%$ of the total coincidences with a $\sim 4\sigma$ confidence level ) should be associated with a probability of chance coincidence of 1.8$\times 10^{-4}$. Similarly, for displacing the \textit{Fermi} sources in 2$^\circ$-bins, we obtain an average of coincidences of 3.6$\pm$1.8. That result indicates that 8.4 of the 12 coincident sources ($\sim 70\%$ of the total coincidences with a $\sim 5\sigma$ confidence level) should be associated with a probability of $5.6 \times 10^{-6}$ of being chance coincidences. In a similar way, the association of \textit{Fermi} sources with WR and Of-type stars is unclear since the number of actual coincidences and the results of the Monte Carlo simulations are too similar and thus the probability of chance coincidence is too high ($0.29$ and $0.11$ for WR stars and Of-type stars, respectively). The results for WR and Of-type stars are different from those obtained by \cite{rome99} for \textit{EGRET} sources. The probability of chance coincidences has increased while the number of candidates decreased in the case of WR stars. In the case of Of-type stars the probability of chance association has increased as the number of coincidences increased as well. The probability of chance coincidence with OB associations is as low as $\sim 10^{-6}$. In this case, however, the nature of the gamma-ray emission is not clear, as it is discussed in section \ref{sec:obasso}

\subsection{Young stellar objects}{\label{subsec:ysos}}
The association of gamma-ray sources and massive YSOs has been suggested in the last years after studying a reasonable scenario for the production of non~thermal emission \citep{ara07}. However, this is the first time that the study of YSOs as gamma-ray sources is carried out in a statistical way, taking advantage of the \textit{Fermi} catalog. 

The results of our cross-check, shown in Table \ref{tab:yso}, indicate that 12 gamma-ray sources are positionally coincident with 23 YSOs. In this table we present, from left to right, the \textit{Fermi} source name, its J2000 equatorial coordinates, its positional uncertainty, the spectral $\gamma$-ray index, the energy flux (E$>$100 MeV), the YSO name, its J2000 equatorial coordinates, the angular distance between the two compared sources, the distance to the YSO, its IR luminosity, and the mass of the star forming region where it is embedded.

In what follows we present a case by case discussion of the gamma-ray fields.\\

- 1FGL J0541.1+3542. This source is coincident with three YSOs: G173.6328+02.8604, G173.6339+02.8218 and G173.6882+02.7222.  Their luminosities are below 4.8$\times 10^{4}L_\odot$. The three objects belong  to the G173.6036+02.6237 complex (in the RMS Survey notation), which is situated at a distance of 1.6 kpc. We also find two Herbig Haro like objects, GGD 5 and GGD 6 \citep{ggd}, within the error ellipse of the \textit{Fermi} source. Outside the \textit{Fermi} error box the complex also harbors three more YSOs and an HII region.\\

- 1FGL J0647.3+0031. The YSO G212.0641-00.7395 is the only coincidence with this source. Its kinematic distance is 6.4 kpc and it has a luminosity of 2.5$\times 10^4 L_\odot$. This source belongs to the G211.9800-00.9710 complex, together with another YSO that lies outside the error box of the gamma-ray source. \\

-  1FGL J1256.9$-$6337. We find G303.5990-00.6524 within the error ellipse of this gamma-ray source. It is a YSO with a bolometric luminosity of $8.3\times 10^3 L_\odot$ and located at a kinematic distance of 11.3 kpc. It belongs to the G303.5670-00.6253 complex that also harbors an HII region. \\

- 1FGL J1315.0$-$6235. This source is coincident with G305.4840+00.2248, which is a YSO  with a luminosity of $3.8 \times 10^3 L_\odot$ and located at a distance of 3.6 kpc.  \\

- 1FGL J1651.5-4602.  The source G339.8838-01.2588 is coincident with this gamma-ray source. It is a YSO with a bolometric luminosity of 2.1$\times 10^4 L_\odot$. It is located at 2.6 kpc from the Earth. This source has been detected in radio at 8.6 GHz with an integrated flux of 2.6 mJy \citep{walsh98}. There is no indication of whether the radio flux is non~thermal or not.\\

- 1FGL J1702.4$-$4147. This source is coincident with two YSOs: G344.4257+00451B and G344.4257+00451C, which form the G344.4120+00.0492 complex, together with two HII regions. They are located at 5 kpc from the Earth and both have a bolometric luminosity of 1.5$\times 10^4 L_\odot$. These gamma-ray source is also near to a cluster of stars \citep[see][]{dutra03} and a molecular cloud (see \cite{russeil04}). \\

- 1FGL J1846.8$-$0233.  We find a coincidence with the source G030.1981-00.1691. It has a bolometric luminosity of 2.9$\times 10^4 L_\odot$ and is located at 7.4 kpc. Within the error ellipse of the \textit{Fermi} source, there are several other sources, such as dark nebulae and HII regions. \\

- 1FGL J1848.1$-$0145. This source is coincident with two YSOs, G030.9726-00.1410 and G030.9959-00.0771, located at 5.7 kpc and with bolometric luminosities of 3.9$\times 10^3 L_\odot$ and 5.1$\times 10^3 L_\odot$, respectively. Both are part of the G031.1451+00.0383 complex, which hosts five more YSOs (outside the \textit{Fermi} error box), five diffuse HII regions, and ten HII regions. Within the error ellipse of the gamma-ray emission, there is an unidentified very high-energy gamma-ray source, HESS~J1848-018 \citep{chaves08}, which is probably the most suitable very high-energy candidate counterpart to the \textit{Fermi} detection. \\

- 1FGL J1853.1+0032. This source has the biggest error ellipse ($\Delta\theta_{95\%} \sim 0.5^\circ$). For that reason, there is a high number of sources within its location error box, including pulsars, supernova remnants and X-ray sources. The cross-match of the catalogs yields six coincidences. These six YSOs belong to three different complexes with different distance to the Earth: G032.8205-00.3300 and G033.3891+00.1989 belong to the G033.1844-00.0572 complex, located at 5.1 kpc. The luminosities of these two sources are 1.7 and 1.1 $\times 10^3 L_\odot$, respectively; G033.3933+00.0100 and G33.5237+00.0198 belong to the G033.6106+00.0464 complex, located at a distance of 6.8 kpc and the luminosity is 7.9$\times 10^3 L_\odot$ for both sources; G034.0126-00.2832 and G034.0500-00.2977 belong to the G034.0313-00.2904 complex. The distance to these objects is 13.3 kpc and their bolometric luminosities are 3.4$\times 10^4 L_\odot$ and 2.4$\times 10^4L_\odot$, respectively. \\

- 1FGL J1925.0+1720. Two YSOs are coincident with it: G052.2025+00.7217A and G052.2078+00.6890. The bolometric luminosities are 1.5$\times 10^4 L_\odot$ and 2.0 $\times10^4 L_\odot$, respectively. They belong to the G052.2052+00.7053 complex which is located at 10.2 kpc, and hosts also a HII region.\\

- 1FGL J1943.4+2340. There is spatial coincidence with the source G059.7831+00.0648, located at 2.2 kpc. It has a bolometric luminosity of 6.8$\times 10^4 L_\odot$. This YSO has been detected in radio at 8.6 GHz with an integrated flux of 1.0 mJy \citep{sridharan02}. The infrared counterpart of this YSO is IRAS 19410+2336. It was observed by Chandra in 2002, which found hard X-ray emission from a number of sources within this high-mass star-forming region \citep{beut02}. The region has two cores where star formation takes place, with masses of 840 $M_\odot$ and 190 $M_\odot$ \citep{sridharan02,beuther02b}. In the latter paper, it is proposed that the X-ray emission is produced by magnetic reconnection effects between the protostars and their accretion disks. The interaction of several molecular outflows, where the YSO from the RMS survey is located, and the combined effects of the stellar winds make a good scenario that might result in particle acceleration up to relativistic energies.  \\

- 1FGL J2040.0+4157. There is spatial coincidence with G081.5168+00.1926. This YSO is located at 1.7 kpc and shows a bolometric luminosity of 7.04$\times 10^2 L_\odot$. There is a galaxy (2MASX J20395796+4159152) located at 2.3$^{\prime \prime}$ from the position of that YSO.

\subsection{WR and Of-type stars}

We found two \textit{Fermi} sources coincident with 15 WR stars in the Galactic center cluster and nine WR stars in the Quintuplet cluster. Some of these stars \citep[see][]{vander06} show variability and have been detected at X-rays with evidence of nonthermal emission.
The results of our study are shown in Table \ref{tab:wr}.
We cannot state that these two coincidences correspond to physical associations given the high chance association probability obtained from the Monte Carlo study. The first one is in the direction of the galactic center, and there are several other sources that introduce confusion. The second one was suggested as potential association with the Pistol Star in \cite{abdo10} and it is situated in a very crowded field. 

In the case of Of-type stars, which are the evolved state of O stars, and precursors of WR stars, our results show that five \textit{Fermi} sources are coincident with five stars (see Table \ref{tab:of}). The probability of chance coincidence is too high to state a physical association. The sources 1FGL J1315.0-6235 and 1FGL J1853.1+0032 also show positional coincidence with YSOs (see Table \ref{tab:yso}). The case of association with YSOs has a much lower value for the chance probability. The source 1FGL~J1112.1-6041 is coincident with HD~97434, a multiple star system. 1FGL~J1315.0-6235 is located in a regions that harbors dark nebulae and molecular clouds. The Of star coincident with this source is HD~115071, which is a spectroscopic binary. Finally, 1FGL J2004.7+3343 is coincident with HD~227465, and there is also the source G70.7+1.2 inside the error box of the \textit{Fermi} detection, which might contain a Be star and an X-ray-emitting B star pulsar binary \citep{cameron07}.

\begin{small}
\begin{table*}[h!]
  \begin{center}
  \caption{Positional coincidence of \textit{Fermi} sources with WR stars.\label{tab:wr}}    
    \scalebox{0.92}[0.92]{
    \begin{tabular}{llllr@{$\pm$}lr@{$\pm$}llllll}
    \hline
    \hline
Fermi Name&RA&Dec&95\% Semi& \multicolumn{2}{c}{$\Gamma$}& \multicolumn{2}{c}{Flux(E$>$100 MeV)}&Star&RA&Dec&$\Delta\theta$&Distance\\
(1FGL)&$\left(^\circ \right)$&$\left(^\circ \right)$&Major Axis ($^\circ$)& \multicolumn{2}{c}{$(F \propto E^{-\Gamma})$}& \multicolumn{2}{c}{$\times 10^{-10}$ erg cm$^{-2}$ s$^{-1}$}&&$\left(^\circ \right)$&$\left(^\circ \right)$&$\left(^\circ \right)$&$\left( \rm{kpc}\right)$\\
\hline
J1745.6-2900 &266.420&$-$29.014&0.019&2.26&0.03&7.1&0.7&WR 101a\tablefootmark{1}&266.41454&$-$29.00950&0.01&8.0 \\ \\
J1746.4-2849 &266.618&$-$28.818&0.10&2.2&0.5&4.6&0.0&WR 102c\tablefootmark{2}&266.54667&$-$28.81822&0.06&8.0\\
\hline
 \end{tabular}}
 \tablefoot{\tablefoottext{1}{WR stars 101 b, c, d, e, f, g, h, i, j, k, l, m, n, o, belonging to the Galactic Center Cluster lie at distances less than $20^{\prime\prime}$ and are also within the error box of the gamma-ray source;}
 \tablefoottext{2}{WR stars 102 d, e, f, g, h, i, j, k belonging to the Quintuplet Cluster lie at distances less than $1^{\prime}$ and are also coincident with the gamma-ray source.}}
  \end{center}
\end{table*}
\end{small}

\subsection{OB associations}\label{sec:obasso}

Our study yields 107 \textit{Fermi} sources positionally coincident with 35 OB associations, which represent $\sim 41\%$ of the sample. We list the results in Table \ref{tab:obassoc}, available in the electronic version. We get this large number of gamma-ray sources because of the size of the OB associations. Most of them have angular sizes of $\sim 1^\circ$ in diameter or higher. Our results extend those of \cite{rome99}, where they found 26 coincidences between \textit{EGRET} sources and OB associations. Most of the OB associations that they found are present in our results, along with other new candidates, as expected from the higher sensitivity of \textit{LAT}.  Although the number of sources has increased, the probability of chance associations is approximately the same as in \cite{rome99}.

Most of the OB associations have fewer than five \textit{Fermi} sources within their error boxes. There are seven OB associations with five or more \textit{Fermi} sources within their error boxes. There are four associations with a number of gamma-ray coincidences between five and ten (Ori 1 B, Ori 1 C, Car 2, and Cyg 1,8,9), located at short distances from the Earth (less than 1 kpc) and located at galactic latitudes of $\sim |15^\circ|$ (association centroid position). There is an exception, Car 2, which is located at 2.2 kpc and has a galactic latitude of $-0.13^\circ$.  The OB associations with the large number of gamma-ray coincidences ($> 10$) are those from Scorpius (Sco 2 A, Sco 2 B and Sco 2 D). Those associations are located at $\sim$ 170 pc on average and have very large angular sizes ($\sim 9.5^\circ$ on average). 

There are five \textit{Fermi} sources that are coincident with OB associations and YSOs at the same time: 1FGL J1256.9-6337, 1FGL~J1315.5-6235, 1FGL~J1702.4-447, 1FGL~J1943.4+2340, and 1FGL~J2040.0+4157. In all cases the OB association overlaps both the \textit{Fermi} source and the YSO. In all cases but one, however, the distances to the YSO and the OB association are much too different.

\begin{small}
\begin{table*}[h!]

  \begin{center}
  \caption{Positional coincidence of \textit{Fermi} sources with Of stars.\label{tab:of}}   
    \scalebox{0.92}[0.92]{
    \begin{tabular}{llllr@{$\pm$}lr@{$\pm$}llllll}
    \hline
    \hline
Fermi Name&RA&Dec&95\% Semi& \multicolumn{2}{c}{$\Gamma$}& \multicolumn{2}{c}{Flux(E$>$100 MeV)}&Star&RA&Dec&$\Delta\theta$&Distance\tablefootmark{1}\\
(1FGL)&$\left(^\circ \right)$&$\left(^\circ \right)$&Major Axis ($^\circ$)& \multicolumn{2}{c}{$(F \propto E^{-\Gamma})$}& \multicolumn{2}{c}{$\times 10^{-12}$ erg cm$^{-2}$ s$^{-1}$}&&$\left(^\circ \right)$&$\left(^\circ \right)$&$\left(^\circ \right)$&$\left(\rm{kpc}\right)$\\
\hline
J0005.1+6829  &	  1.2841 &  68.4883   & 0.4427 & 2.58 & 0.12 & 1.7  & 0.5  & BD+67 1598 &   1.225 &  68.167   & 0.32 & 1.07\\
J1112.1-6041  & 168.0486 & $-$60.6929 & 0.0461 & 2.12 & 0.05 & 14.1 & 1.6  & HD 97434   & 167.975 & $-$60.683 & 0.04 & 2.67\\
J1315.0-6235  & 198.7644 & $-$62.5971 & 0.1860 & 2.31 & 0.12 & 6.9  & 0.0  & HD 115071  & 199.000 & $-$62.583 & 0.11 & 0.74\\
J1853.1+0032  & 283.2887 &   0.5369   & 0.5207 & 2.18 & 0.07 & 5.7  & 1.7  & BD-0 3584  & 283.400 &   0.567   & 0.12 & 2.18\\
J2004.7+3343  & 301.1855 &  33.7171   & 0.1320 & 2.28 & 0.08 & 5.2  & 0.8  & HD 227465  & 301.125 &  33.700   & 0.05 & 3.48\\
\hline

 \hline
 \end{tabular}
 }
 \tablefoot{\tablefoottext{1}{distances from \cite{cruz74}}}
  \end{center}
\end{table*}
\end{small}

\section{Discussion}

\begin{figure}
\resizebox{\hsize}{!}{\includegraphics{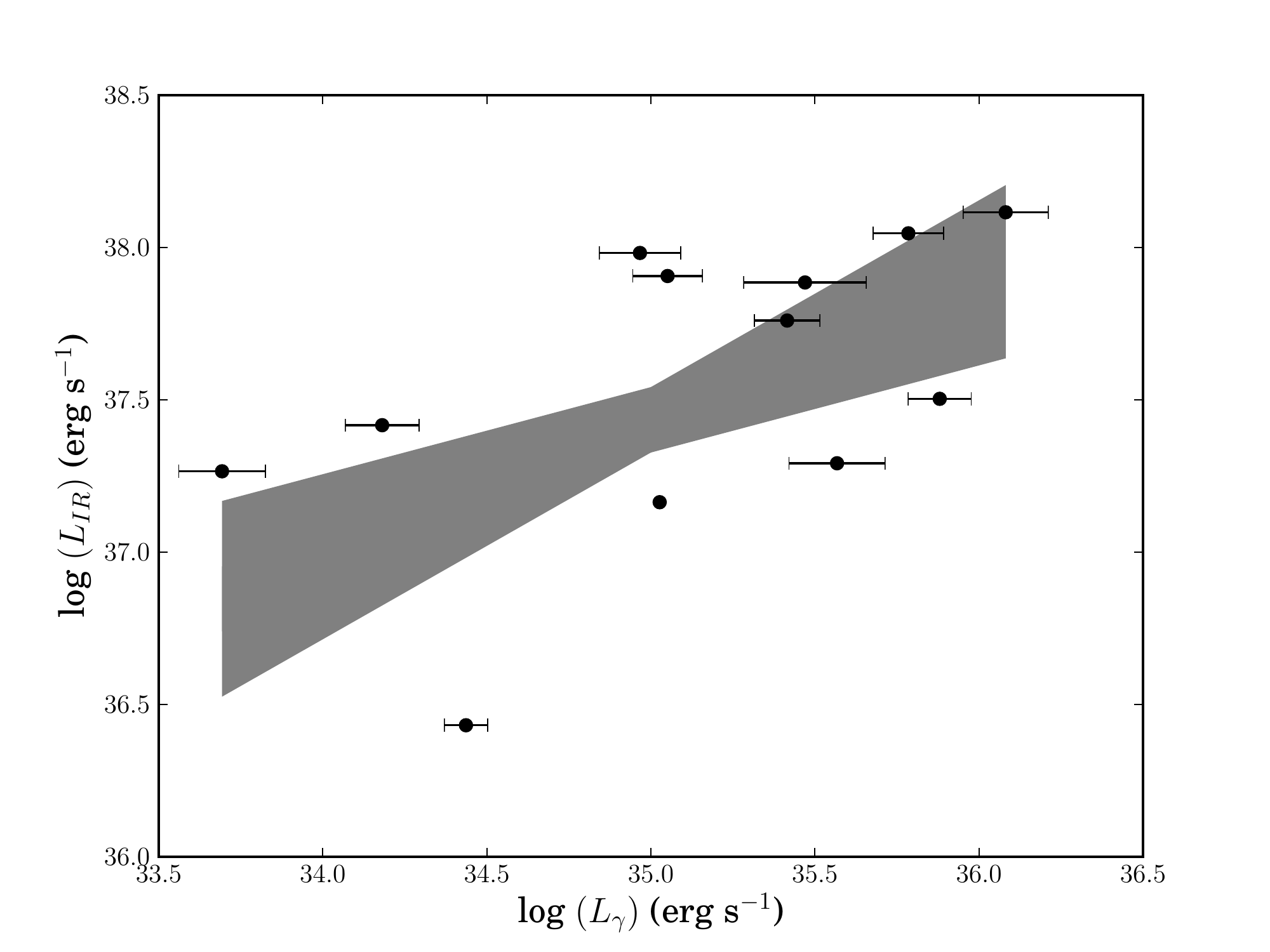}}
  \caption{IR luminosity versus gamma-ray luminosity above 100 MeV of the brightest YSOs coincident with each \textit{Fermi} source. The shaded area represents the 1-$\sigma$ confidence interval of a least square fit taking eight data points randomly and repeating the fit one thousand times. It can be roughly  seen that the higher the gamma-ray luminosity, the higher the IR luminosity. \label{fig:lumi}}
\end{figure}	

The fact that there are five \textit{Fermi} sources coincident with both YSOs and OB associations makes us consider which is the chance probability of having coincidence of \textit{Fermi} sources with YSOs alone. Using the Monte Carlo algorithm again, but taking this constrain into account, we obtain a mean value of coincidences of $2.8\pm 1.7$, and the chance probability for the seven \textit{Fermi} sources coincident only with YSOs is $\sim 1.6\%$.

The brightest IR YSOs have more molecular mass available for proton-proton collisions and Bremsstrahlung interactions than those that are faint. It is expected then that the brightest IR YSOs would show the highest gamma-ray luminosity. To test such a trend, we have plotted in Figure \ref{fig:lumi} the IR luminosity versus the gamma-ray luminosity of the brightest YSOs coincident with each \textit{Fermi} source. The gamma-ray luminosity was calculated from the gamma-ray flux (see Table \ref{tab:yso}) assuming that the distance to the gamma-ray source is equal to the distance to the corresponding YSO. We see that, under this assumption, there is a trend toward increasing the gamma-ray luminosity as the IR luminosity increases, too. This trend, however, is obtained using the whole data set, and according to the results from our Monte Carlo simulations, we should have four chance coincidences out of the 12 candidates. To see how this could affect the trend, we selected eight data points randomly and fitted them to a straight line by least squares. We repeated this process one thousand times, getting an average least square fit within one standard deviation. The limits of this fit are plotted in Figure 2. The increasing trend in the data is visible, although with a significant dispersion. This is not surprising considering the broad approach.

The association of WR stars with gamma-ray sources has been discussed in the past by \cite{kaul97} and \cite{rome99}. In both cases, the authors studied the positional coincidence between WR stars and unidentified EGRET sources. In the later work, they find that two WR stars are of special interest: WR 140 and WR 142. The first one is a binary system, composed of a WC 7 plus an O4-5 star where the region of collision of the winds seems to be a good place for particle acceleration and high energy emission. It should be mentioned, however, that WR 140 is a long-period binary with variability expected on time scales of years \citep{williams87}. In the second one, WR 142, the hard X-ray emission and fast wind may indicate a colliding wind shock that could be explained by a companion close to this star \citep{sokal10}. No companion, however, has been reported so far.
In our work none of these mentioned WR stars appear to be coincident with any \textit{Fermi} source. The poor statistical correlations found from the simulations does not allow us to be confident with any of the coincidences found. The case of Of-type stars is also unclear since the probability of chance association is high. 

Finally, our results for the OB associations are not conclusive. We list the results in Table \ref{tab:obassoc}, available in the electronic version. The probability of chance coincidence is negligible ($\sim 10^{-6}$), but we do get several gamma-ray sources for each OB association. Thus, is very difficult to assign a specific counterpart to the gamma-ray emission.

\section{Conclusions}

We studied the two-dimensional coincidence between unidentified \textit{Fermi} sources and catalogs of galactic young objects, such as YSOs, WR stars, Of stars, and OB associations. We have found a statistical correlation between gamma-ray sources and YSOs. The correlation with the early type stars with strong winds remain unclear, since the candidates are located in crowded fields with many other alternatives to the gamma-ray emission, and the probability of chance associations is high. In the case of OB associations, the probability of chance associations is negligible. However, we cannot assign a specific counterpart to the gamma-ray emission because of the high angular size of most of OB associations.

What we have presented here is the first statistical evidence for gamma-ray emission from massive YSOs.

\begin{acknowledgements} 
The authors acknowledge support from the Spanish Ministerio de Ciencia e Innovaci\'on (MICINN) under grant AYA2010-21782-C03-01. G.E.R. also acknowledges support of PIP 0078 (CONICET).
\end{acknowledgements}

\bibliography{biblio}{}
\bibliographystyle{aa}

\clearpage
\onecolumn
\begin{landscape}
      \begin{small}
 \begin{tiny}
  \begin{center}
     
    \begin{longtable}{llllr@{$\pm$}lr@{$\pm$}lllllll}
          \caption{Positional coincidence of \textit{Fermi} sources with OB associations.\label{tab:obassoc}}\\
    \hline
     \hline
Fermi Name&RA&Dec& 95\% Semi & \multicolumn{2}{c}{$\Gamma$} & \multicolumn{2}{c}{Flux(E$>$100 MeV)} & OB & RA & Dec & Distance\tablefootmark{1} & $\Delta\theta$ & OB Size\tablefootmark{1} \\
(1FGL)&$\left(^\circ \right)$&$\left(^\circ \right)$&Major Axis ($^\circ$)& \multicolumn{2}{c}{$(F \propto E^{-\Gamma})$}& \multicolumn{2}{c}{$\times 10^{-12}$ erg cm$^{-2}$s$^{-1}$}&association&$\left(^\circ \right)$&$\left(^\circ \right)$&$\left(\rm{kpc}\right)$&$\left(^\circ \right)$&$\left(^\circ \right)$ \\
\hline
\hline
\endhead

\hline \multicolumn{14}{r}{{Continued on next page}} \\ \hline
\endfoot

J0205.6+6449   &31.4090  &64.8286     &    --  & 2.36 & 0.06 & 3.78 & 0.60 &CLUST  1 & 31.03140  & 64.70406&0.89  &0.20&   0.9 \\ 
J0214.1+6020   &33.5476  &60.3446     & 0.1624 & 2.1 & 0.2 & 1.95 & 0.68   &PER 1  B & 33.47470  & 59.59772&1.60  &0.75&   0.8 \\ 
J0521.6+0103   &80.4230  & 1.0512     & 0.1077 & 2.1 & 0.2 & 0.92 & 0	   &ORI 1  B & 83.30102  & -0.39051&0.50  &3.22&   6.6 \\ 
J0534.7-0531   &83.6872  &-5.5222     & 0.1035 & 2.37 & 0.09 & 3.51 & 0.58 &ORI 1  B & 83.30102  & -0.39051&0.50  &5.15&   6.6 \\ 
   		 &&&&\multicolumn{2}{c}{}&\multicolumn{2}{c}{}  	   &ORI 1  C & 83.86341  & -5.53543&0.66  &0.18&   3.0 \\ 
J0536.2-0607   &84.0623  &-6.1285     & 0.0857 & 2.42 & 0.07 & 3.79 & 0.40 &ORI 1  B & 83.30102  & -0.39051&0.50  &5.79&   6.6 \\ 
	     &&&&\multicolumn{2}{c}{}&\multicolumn{2}{c}{}		   &ORI 1  C & 83.86341  & -5.53543&0.66  &0.63&   3.0 \\ 
J0539.4-0400   &84.8703  &-4.0166     & 0.2809 & 2.5 & 0.1 & 1.06 & 0.45   &ORI 1  B & 83.30102  & -0.39051&0.50  &3.95&   6.6 \\ 
	 &&&&\multicolumn{2}{c}{}&\multicolumn{2}{c}{}  		   &ORI 1  C & 83.86341  & -5.53543&0.66  &1.82&   3.0 \\ 
J0540.4-0737   &85.1220  &-7.6268     & 0.0999 & 2.2 & 0.1 & 1.31 & 0.45   &ORI 1  C & 83.86341  & -5.53543&0.66  &2.44&   3.0 \\ 
J0540.9-0547   &85.2317  &-5.7900     & 0.3870 & 2.4 & 0.1 & 1.29 & 0.45   &ORI 1  B & 83.30102  & -0.39051&0.50  &5.73&   6.6 \\ 
	       &&&&\multicolumn{2}{c}{}&\multicolumn{2}{c}{}		   &ORI 1  C & 83.86341  & -5.53543&0.66  &1.39&   3.0 \\ 
J0541.9-0204   &85.4885  &-2.0754     & 0.0791 & 2.3 & 0.1 & 1.77 & 0.47   &ORI 1  B & 83.30102  & -0.39051&0.50  &2.76&   6.6 \\ 
J0547.0+0020   &86.7698  & 0.3362     & 0.1868 & 2.50 & 0.09 & 1.61 & 0.43 &ORI 1  B & 83.30102  & -0.39051&0.50  &3.54&   6.6 \\ 
J0636.0+0458   &99.0050  & 4.9806     & 0.0724 & 2.3 & 0.1 & 4.76 & 0.75   &MON 1  B & 99.37146  &  4.81144&1.48  &0.40&   0.6 \\ 
J0705.9-1051  &106.4985  &-10.8528    & 0.1715 & 2.5 & 0.1 & 1.72 & 0.75   &CMA 1  A &106.03646  &-11.01884&0.75  &0.48&   1.0 \\ 
J0709.0-1116  &107.2576  &-11.2758    & 0.0942 & 1.3 & 0.6 & 2.08 & 0.66   &CMA 1  B &107.10737  &-12.28821&1.30  &1.02&   1.2 \\ 
J0841.9-4620  &130.4865  &-46.3448    & 0.2748 & 1.99 & 0.09 & 2.42 & 0.68 &VELA 1 A &129.95838  &-45.96976&1.43  &0.52&   1.1 \\ 
J0854.0-4632  &133.5039  &-46.5424    & 0.1569 & 2.10 & 0.09 & 4.19 & 1.23 &VELA 1 B &132.76816  &-45.93533&1.41  &0.79&   1.0 \\ 
J1045.2-5942  &161.3059  &-59.7059    & 0.0231 & 2.14 & 0.03 & 26.2 & 1.60 &CAR 1  E &161.24112  &-59.73526&2.64  &0.04&   1.5 \\ 
J1057.2-6026  &164.3242  &-60.4495    & 0.0779 & 2.25 & 0.06 & 6.86 & 1.43 &CAR 1-2  &164.67677  &-60.20707&2.62  &0.30&   0.5 \\ 
	       &&&&\multicolumn{2}{c}{}&\multicolumn{2}{c}{}		   &CAR 2    &167.12384  &-60.51125&2.16  &1.38&   2.5 \\ 
J1104.0-6047  &166.0209  &-60.7974    & 0.0633 & 2.27 & 0.06 & 6.84 & 2.59 &CAR 2    &167.12384  &-60.51125&2.16  &0.61&   2.5 \\ 
J1106.7-6150  &166.6808  &-61.8449    & 0.0649 & 2.3 & 0.1 & 5.46 & 1.29   &CAR 2    &167.12384  &-60.51125&2.16  &1.35&   2.5 \\ 
J1112.1-6041  &168.0486  &-60.6929    & 0.0461 & 2.12 & 0.05 & 14.2 & 1.60 &CAR 2    &167.12384  &-60.51125&2.16  &0.49&   2.5 \\ 
J1115.2-6124  &168.8028  &-61.4022    & 0.1037 & 2.25 & 0.08 & 9.73 & 0    &CAR 2    &167.12384  &-60.51125&2.16  &1.21&   2.5 \\ 
J1119.4-6127  &169.8532  &-61.4528    & 0.0576 & 2.15 & 0.08 & 5.51 & 1.66 &CAR 2    &167.12384  &-60.51125&2.16  &1.62&   2.5 \\ 
		&&&&\multicolumn{2}{c}{}&\multicolumn{2}{c}{}		   &NGC 3576 &169.66457  &-61.13154&2.58  &0.33&   0.3 \\ 
J1124.6-5916  &171.1632  &-59.2720    &   --   & 2.36 & 0.06 & 4.54 & 0.79 &CAR 2    &167.12384  &-60.51125&2.16  &2.38&   2.5 \\ 
J1127.7-6244  &171.9345  &-62.7412    & 0.1243 & 2.2 & 0.1 & 3.99 & 1.43   &CRU 1  A &172.81985  &-63.46201&2.58  &0.82&   1.0 \\ 
	    &&&&\multicolumn{2}{c}{}&\multicolumn{2}{c}{}		   &SCO 2  A &188.20526  &-61.74213&0.16  &7.64&   8.0 \\ 
J1134.8-6055  &173.7198  &-60.9320    & 0.0634 & 2.35 & 0.07 & 4.16 & 1.01 &SCO 2  A &188.20526  &-61.74213&0.16  &7.00&   8.0 \\ 
J1136.0-6226  &174.0206  &-62.4342    & 0.1073 & 2.2 & 0.1 & 4.16 & 1.09   &SCO 2  A &188.20526  &-61.74213&0.16  &6.68&   8.0 \\ 
J1207.4-6239  &181.8754  &-62.6552    & 0.0719 & 2.35 & 0.06 & 8.20 & 1.08 &SCO 2  A &188.20526  &-61.74213&0.16  &3.09&   8.0 \\ 
J1213.7-6240  &183.4475  &-62.6743    & 0.0988 & 2.29 & 0.08 & 4.77 & 1.34 &SCO 2  A &188.20526  &-61.74213&0.16  &2.41&   8.0 \\ 
J1234.0-5736  &188.5059  &-57.6129    & 0.0826 & 2.0 & 0.1 & 1.26 & 0.53   &SCO 2  A &188.20526  &-61.74213&0.16  &4.13&   8.0 \\ 
J1241.6-6240  &190.4160  &-62.6798    & 0.1885 & 2.51 & 0.09 & 4.71 & 1.04 &SCO 2  A &188.20526  &-61.74213&0.16  &1.39&   8.0 \\ 
J1256.1-5922  &194.0496  &-59.3778    & 0.1363 & 2.2 & 0.2 & 1.19 & 0.50   &SCO 2  A &188.20526  &-61.74213&0.16  &3.72&   8.0 \\ 
J1256.9-6337  &194.2482  &-63.6213    & 0.1954 & 2.2 & 0.1 & 4.97 & 1.13   &SCO 2  A &188.20526  &-61.74213&0.16  &3.35&   8.0 \\ 
J1300.7-5547  &195.1930  &-55.7989    & 0.2962 & 2.8 & 0.1 & 1.92 & 0.49   &SCO 2  A &188.20526  &-61.74213&0.16  &6.96&   8.0 \\ 
J1301.4-6245  &195.3587  &-62.7609    & 0.1266 & 2.29 & 0.09 & 5.22 & 1.94 &SCO 2  A &188.20526  &-61.74213&0.16  &3.48&   8.0 \\ 
J1306.4-6038  &196.6202  &-60.6381    & 0.1336 & 2.2 & 0.1 & 1.76 & 0.67   &SCO 2  A &188.20526  &-61.74213&0.16  &4.20&   8.0 \\ 
J1307.3-6701  &196.8482  &-67.0235    & 0.1657 & 2.6 & 0.1 & 1.32 & 0.50   &SCO 2  A &188.20526  &-61.74213&0.16  &6.47&   8.0 \\ 
J1309.9-6229  &197.4964  &-62.4869    & 0.1471 & 2.2 & 0.1 & 5.16 & 1.48   &SCO 2  A &188.20526  &-61.74213&0.16  &4.41&   8.0 \\ 
J1315.0-6235  &198.7644  &-62.5971    & 0.1860 & 2.3 & 0.1 & 6.86 & 0	   &SCO 2  A &188.20526  &-61.74213&0.16  &5.00&   8.0 \\ 
J1317.5-6318  &199.3793  &-63.3037    & 0.0685 & 2.1 & 0.1 & 6.64 & 0	   &SCO 2  A &188.20526  &-61.74213&0.16  &5.39&   8.0 \\ 
J1320.6-6258  &200.1639  &-62.9693    & 0.1976 & 2.3 & 0.1 & 4.60 & 1.42   &SCO 2  A &188.20526  &-61.74213&0.16  &5.68&   8.0 \\ 
J1322.0-4515  &200.5154  &-45.2652    & 0.2206 & 2.8 & 0.1 & 4.01 & 0.52   &SCO 2  B &211.14639  &-44.22469&0.18  &7.62&   7.8 \\ 
J1325.6-4300  &201.4148  &-43.0110    & 0.0796 & 2.71 & 0.06 & 7.94 & 0.58 &SCO 2  B &211.14639  &-44.22469&0.18  &7.15&   7.8 \\ 
J1328.2-4729  &202.0508  &-47.4991    & 0.0986 & 2.1 & 0.1 & 2.53 & 0.51   &SCO 2  B &211.14639  &-44.22469&0.18  &7.13&   7.8 \\ 
J1333.4-4036  &203.3574  &-40.6042    & 0.4217 & 2.7 & 0.1 & 1.80 & 0.40   &SCO 2  B &211.14639  &-44.22469&0.18  &6.80&   7.8 \\ 
J1334.2-4448  &203.5635  &-44.8136    & 0.1873 & 2.4 & 0.1 & 1.23 & 0.46   &SCO 2  B &211.14639  &-44.22469&0.18  &5.44&   7.8 \\ 
J1347.8-3751  &206.9674  &-37.8551    & 0.1517 & 2.7 & 0.1 & 2.13 & 0.39   &SCO 2  B &211.14639  &-44.22469&0.18  &7.11&   7.8 \\ 
J1400.1-3743  &210.0356  &-37.7177    & 0.1705 & 1.9 & 0.2 & 1.67 & 0	   &SCO 2  B &211.14639  &-44.22469&0.18  &6.56&   7.8 \\ 
J1407.5-4256  &211.8885  &-42.9485    & 0.1592 & 2.1 & 0.2 & 1.06 & 0.39   &SCO 2  B &211.14639  &-44.22469&0.18  &1.38&   7.8 \\ 
J1417.7-4407  &214.4286  &-44.1320    & 0.2160 & 2.3 & 0.2 & 1.98 & 0.45   &SCO 2  B &211.14639  &-44.22469&0.18  &2.36&   7.8 \\ 
J1417.7-5030  &214.4405  &-50.5111    & 0.2019 & 2.6 & 0.2 & 1.57 & 0.50   &SCO 2  B &211.14639  &-44.22469&0.18  &6.67&   7.8 \\ 
J1428.2-4204  &217.0711  &-42.0711    & 0.0982 & 2.31 & 0.07 & 2.99 & 0.42 &SCO 2  B &211.14639  &-44.22469&0.18  &4.83&   7.8 \\ 
J1501.6-4204  &225.4096  &-42.0819    & 0.4820 & 2.9 & 0.2 & 1.62 & 0.42   &SCO 2  C &228.91604  &-42.39408&0.16  &2.61&   5.8 \\ 
J1513.2-5904  &228.3209  &-59.0821    & 0.0723 & 1.6 & 0.3 & 6.56 & 0	   &PIS 20   &229.83477  &-60.14507&0.96  &1.31&   1.3 \\ 
J1514.1-4745  &228.5328  &-47.7527    & 0.1479 & 2.3 & 0.1 & 1.45 & 0.49   &SCO 2  C &228.91604  &-42.39408&0.16  &5.37&   5.8 \\ 
J1514.7-5917  &228.6808  &-59.2924    & 0.1121 & 1.6 & 0.3 & 4.40 & 0	   &PIS 20   &229.83477  &-60.14507&0.96  &1.03&   1.3 \\ 
J1542.9-2559  &235.7479  &-25.9849    & 0.1772 & 2.4 & 0.2 & 1.21 & 0.39   &SCO 2  D &243.54161  &-23.91278&0.18  &7.36&  11.7 \\ 
J1548.7-2250  &237.1930  &-22.8394    & 0.0732 & 2.0 & 0.2 & 1.92 & 0.47   &SCO 2  D &243.54161  &-23.91278&0.18  &5.93&  11.7 \\ 
J1548.9-5509  &237.2377  &-55.1647    & 0.0942 & 2.36 & 0.09 & 3.63 & 0.11 &NOR 1    &238.67814  &-54.75670&1.10  &0.92&   1.1 \\ 
J1553.4-2425  &238.3711  &-24.4210    & 0.1813 & 2.4 & 0.2 & 1.30 & 0.44   &SCO 2  D &243.54161  &-23.91278&0.18  &4.74&  11.7 \\ 
J1553.5-3116  &238.3877  &-31.2713    & 0.0718 & 1.7 & 0.2 & 1.17 & 0.39   &SCO 2  D &243.54161  &-23.91278&0.18  &8.66&  11.7 \\ 
J1554.0-5345  &238.5184  &-53.7663    & 0.1074 & 2.24 & 0.07 & 15.6 & 1.63 &NOR 1    &238.67814  &-54.75670&1.10  &0.99&   1.1 \\ 
J1600.7-3055  &240.1792  &-30.9264    & 0.0973 & 1.8 & 0.1 & 1.25 & 0	   &SCO 2  D &243.54161  &-23.91278&0.18  &7.62&  11.7 \\ 
J1607.5-2030  &241.9000  &-20.5118    & 0.0681 & 2.3 & 0.2 & 1.29 & 0.42   &SCO 2  D &243.54161  &-23.91278&0.18  &3.73&  11.7 \\ 
J1613.6-5100  &243.4201  &-51.0010    & 0.0661 & 2.20 & 0.06 & 7.65 & 2.34 &R 103  B &243.67674  &-51.07046&3.00  &0.18&   0.3 \\ 
J1614.7-5138  &243.6767  &-51.6411    & 0.1434 & 2.15 & 0.06 & 12.8 & 1.91 &R 103  A &244.40278  &-51.81465&3.22  &0.48&   0.4 \\ 
J1620.9-2731  &245.2310  &-27.5272    & 0.1896 & 2.4 & 0.1 & 1.49 & 0.48   &SCO 2  D &243.54161  &-23.91278&0.18  &3.92&  11.7 \\ 
J1623.5-2345  &245.8923  &-23.7527    & 0.3570 & 2.3 & 0.1 & 3.62 & 0	   &SCO 2  D &243.54161  &-23.91278&0.18  &2.16&  11.7 \\ 
J1625.7-2524  &246.4282  &-25.4163    & 0.0677 & 2.36 & 0.06 & 4.52 & 1.02 &SCO 2  D &243.54161  &-23.91278&0.18  &3.02&  11.7 \\ 
J1625.8-2429  &246.4749  &-24.4971    & 0.0888 & 2.25 & 0.07 & 7.00 & 1.02 &SCO 2  D &243.54161  &-23.91278&0.18  &2.74&  11.7 \\ 
J1626.2-2956  &246.5667  &-29.9409    & 0.1656 & 2.4 & 0.1 & 1.85 & 0.47   &SCO 2  D &243.54161  &-23.91278&0.18  &6.60&  11.7 \\ 
J1626.2-2038  &246.5702  &-20.6485    & 0.4539 & 2.5 & 0.1 & 1.83 & 0.49   &SCO 2  D &243.54161  &-23.91278&0.18  &4.30&  11.7 \\ 
J1627.8-3204  &246.9511  &-32.0761    & 0.1646 & 2.1 & 0.1 & 1.22 & 0.37   &SCO 2  D &243.54161  &-23.91278&0.18  &8.70&  11.7 \\ 
J1627.8-1711  &246.9712  &-17.1897    & 0.1481 & 2.2 & 0.2 & 0.99 & 0.41   &SCO 2  D &243.54161  &-23.91278&0.18  &7.45&  11.7 \\ 
J1628.6-2419  &247.1638  &-24.3296    & 0.1817 & 2.1 & 0.1 & 4.19 & 0	   &SCO 2  D &243.54161  &-23.91278&0.18  &3.33&  11.7 \\ 
J1632.7-2431  &248.1931  &-24.5203    & 0.1713 & 2.3 & 0.1 & 1.79 & 0.63   &SCO 2  D &243.54161  &-23.91278&0.18  &4.29&  11.7 \\ 
J1632.9-4802  &248.2264  &-48.0402    & 0.0527 & 2.17 & 0.05 & 14.0 & 2.85 &ARA 1A A &249.78617  &-48.95281&1.59  &1.38&   1.3 \\ 
J1640.8-4634  &250.2024  &-46.5816    & 0.0969 & 2.32 & 0.04 & 19.2 & 3.08 &NGC 6204 &251.38377  &-47.30858&1.94  &1.09&   1.5 \\ 
J1645.0-2155  &251.2549  &-21.9319    & 0.1919 & 2.3 & 0.1 & 1.43 & 0.43   &SCO 2  D &243.54161  &-23.91278&0.18  &7.38&  11.7 \\ 
J1648.4-4609  &252.1112  &-46.1599    & 0.0553 & 2.29 & 0.05 & 15.9 & 3.98 &NGC 6204 &251.38377  &-47.30858&1.94  &1.25&   1.5 \\ 
J1702.4-4147  &255.6043  &-41.7857    & 0.1040 & 2.50 & 0.07 & 4.07 & 0.53 &SCO 1    &253.59121  &-41.50018&1.92  &1.53&   1.9 \\ 
J1715.2-3319  &258.8207  &-33.3323    & 0.1397 & 2.3 & 0.1 & 3.11 & 0	   &SCO 4    &258.66248  &-33.68906&1.23  &0.38&   1.2 \\ 
J1717.9-3343  &259.4985  &-33.7269    & 0.0832 & 2.42 & 0.05 & 6.88 & 1.03 &SCO 4    &258.66248  &-33.68906&1.23  &0.70&   1.2 \\ 
J1732.3-3243  &263.0869  &-32.7293    & 0.1243 & 2.33 & 0.05 & 13.9 & 1.91 &TR 27    &263.17260  &-32.92243&1.31  &0.21&   1.1 \\ 
J1810.9-1905  &272.7486  &-19.0843    & 0.0878 & 2.26 & 0.06 & 11.0 & 1.42 &SGR 4    &273.93407  &-18.92533&1.85  &1.13&   1.5 \\ 
J1811.3-1959  &272.8294  &-19.9920    & 0.0985 & 2.1 & 0.2 & 5.75 & 0	   &SGR 7    &273.41595  &-20.64018&1.28  &0.85&   0.8 \\ 
		       &&&&\multicolumn{2}{c}{}&\multicolumn{2}{c}{}	   &SGR 4    &273.93407  &-18.92533&1.85  &1.49&   1.5 \\ 
J1814.0-1736  &273.5201  &-17.6008    & 0.0470 & 2.34 & 0.04 & 14.8 & 1.69 &SGR 4    &273.93407  &-18.92533&1.85  &1.38&   1.5 \\ 
J1817.6-1651  &274.4099  &-16.8602    & 0.1466 & 2.30 & 0.06 & 5.91 & 2.77 &SER 1  A &275.20333  &-16.63491&1.50  &0.79&   0.9 \\ 
J1818.7-1557  &274.6776  &-15.9591    & 0.1114 & 2.34 & 0.07 & 12.4 & 1.92 &SER 1  A &275.20333  &-16.63491&1.50  &0.84&   0.9 \\ 
J1821.1-1425  &275.2907  &-14.4198    & 0.0837 & 2.26 & 0.06 & 7.58 & 2.34 &SCT 3    &276.31989  &-14.33074&1.48  &1.00&   1.1 \\ 
J1823.2-1336  &275.8135  &-13.6060    & 0.0738 & 2.12 & 0.05 & 16.3 & 2.77 &SCT 3    &276.31989  &-14.33074&1.48  &0.88&   1.1 \\ 
J1825.7-1410  &276.4394  &-14.1818    & 0.0553 & 1.1 & 0.3 & 6.86 & 2.29   &SCT 3    &276.31989  &-14.33074&1.48  &0.19&   1.1 \\ 
J1826.2-1450  &276.5630  &-14.8481    &   -    & 2.38 & 0.03 & 28.7 & 2.31 &SCT 3    &276.31989  &-14.33074&1.48  &0.57&   1.1 \\ 
J1943.4+2340  &295.8670  &23.6818     & 0.1118 & 2.2 & 0.1 & 2.62 & 0.67   &VUL 4    &296.49937  & 24.37678&1.21  &0.90&   1.0 \\ 
J1948.6+2437  &297.1532  &24.6293     & 0.1326 & 2.2 & 0.1 & 2.26 & 0.68   &VUL 4    &296.49937  & 24.37678&1.21  &0.65&   1.0 \\ 
J2015.7+3708  &303.9291  &37.1402     & 0.0400 & 2.26 & 0.03 & 13.9 & 1.37 &CYG 1,8,9&305.03403  & 38.92113&1.37  &1.98&   4.9 \\ 
J2020.0+4049  &305.0036  &40.8191     & 0.1014 & 2.12 & 0.08 & 8.16 & 3.01 &CYG 1,8,9&305.03403  & 38.92113&1.37  &1.90&   4.9 \\ 
J2021.5+4026  &305.3817  &40.4460     &   -    & 2.25 & 0.01 & 97.2 & 3.48 &CYG 1,8,9&305.03403  & 38.92113&1.37  &1.55&   4.9 \\ 
J2030.0+3641  &307.5021  &36.6844     & 0.0559 & 2.10 & 0.07 & 3.42 & 1.07 &CYG 1,8,9&305.03403  & 38.92113&1.37  &2.97&   4.9 \\ 
J2032.8+3928  &308.2006  &39.4703     & 0.2507 & 2.59 & 0.07 & 5.12 & 1.39 &CYG 1,8,9&305.03403  & 38.92113&1.37  &2.51&   4.9 \\ 
J2034.7+3639  &308.6968  &36.6512     & 0.1254 & 2.2 & 0.1 & 2.30 & 0.69   &CYG 1,8,9&305.03403  & 38.92113&1.37  &3.68&   4.9 \\ 
J2040.0+4157  &310.0158  &41.9536     & 0.1970 & 2.66 & 0.06 & 7.92 & 1.19 &CYG 1,8,9&305.03403  & 38.92113&1.37  &4.86&   4.9 \\ 
J2207.5+6440  &331.8799  &64.6720     & 0.1971 & 2.65 & 0.08 & 4.14 & 0.52 &CEP 2  B &328.84072  & 61.71559&0.77  &3.26&   3.6 \\ 
J2214.5+5949  &333.6453  &59.8288     & 0.1184 & 2.4 & 0.1 & 2.23 & 0.62   &CEP 2  B &328.84072  & 61.71559&0.77  &3.01    3.6\\ 
J2243.4+4104  &340.8659  &41.0733     & 0.1923 & 2.6 & 0.2 & 1.71 & 0	   &LAC 1    &339.37867  & 39.90053&0.63  &1.63    1.7\\ 
J2250.8+6336  &342.7210  &63.6153     & 0.0687 & 2.33 & 0.07 & 3.39 & 0.43 &CEP 3    &344.37054  & 62.90856&0.84  &1.02    1.9\\  
 
 \hline

 \end{longtable}
 \tablefoot{\tablefoottext{1}{from \cite{cruz74}}}
   \end{center}
  \end{tiny} 
\end{small}
\end{landscape}
\end{document}